\documentstyle[aps,preprint]{revtex}
\input{scrload}

\begin{document}

\tightenlines

\title{Phase diagram of symmetric binary mixtures at equimolar and 
non-equimolar concentrations: a systematic investigation}

\author{D.~Pini}
\address{Istituto Nazionale di Fisica della Materia and Dipartimento di Fisica,
\\ Universit\`a di Milano, Via Celoria 16, 20133 Milano, Italy}

\author{M.~Tau}
\address{Istituto Nazionale di Fisica della Materia and Dipartimento di Fisica,
\\ Universit\`a di Parma, Parco Area delle Scienze 7/A, 43100 Parma, Italy}

\author{A.~Parola}
\address{Istituto Nazionale di Fisica della Materia and Dipartimento di 
Scienze Fisiche, \\
Universit\`a dell'Insubria, Via Valleggio 11, 22100 Como, Italy}

\author{L.~Reatto}
\address{Istituto Nazionale di Fisica della Materia and Dipartimento di Fisica,
\\ Universit\`a di Milano, Via Celoria 16, 20133 Milano, Italy}

\maketitle

\begin{abstract}

We consider symmetric binary mixtures consisting of spherical particles 
with equal diameters interacting via a hard-core plus attractive 
tail potential with strengths $\epsilon_{ij}$, $i,j=1,2$, such that 
$\epsilon_{11}=\epsilon_{22}>\epsilon_{12}$. The phase diagram of the system
at all densities and concentrations
is investigated as a function of the unlike-to-like interaction ratio 
$\delta=\epsilon_{12}/\epsilon_{11}$ by means of the hierarchical reference 
theory (HRT). The results are related to those of previous investigations
performed at equimolar concentration, as well as to the topology of 
the mean-field critical lines. As $\delta$ is increased in the interval 
$0<\delta<1$, we find first a regime where the phase diagram at equal species 
concentration displays a tricritical point, then one where both 
a tricritical and a liquid-vapor critical point are present. We did 
not find any clear evidence of the critical endpoint topology predicted 
by mean-field theory as $\delta$ approaches $1$, at least up to $\delta=0.8$, 
which is the largest value of $\delta$ investigated here.  
Particular attention was paid 
to the description of the critical-plus-tricritical point regime in the whole 
density-concentration plane. In this situation, the phase diagram shows, 
in a certain temperature interval, a coexistence region that encloses 
an island of homogeneous, one-phase fluid.   

\end{abstract} 


\section{Introduction}
\label{sec:introduction}

A major difference between the phase behavior of one-component fluids and 
binary mixtures is that, even if one considers just simple systems 
with a spherically symmetric, Lennard-Jones (LJ) like interaction profile, 
the qualitative features of the phase diagram of mixtures depend very 
sensitively on the parameters of the microscopic potential. If the interaction
between a particle of species $i$ and a particle of species $j$ is modeled 
as the sum of a hard-core repulsion and a longer-ranged attractive tail, 
the relevant parameters are the hard-sphere diameters $\sigma_{ij}$ and the
strengths $\epsilon_{ij}$ of the attractive contributions, $i,j=1,2$. 
This parameter space is drastically reduced by focusing on a particular class
of systems, generally referred to as {\em symmetric mixtures}, such that 
$\sigma_{11}=\sigma_{22}=\sigma_{12}=\sigma$, 
$\epsilon_{11}=\epsilon_{22}=\epsilon$. Since the quantities $\sigma$ 
and $\epsilon$ can be included into the definition of the temperature $T$ and 
number densities $\rho_{i}$ by introducing standard reduced units 
$\rho^{*}_{i}=\rho_{i}\sigma^{3}$, $T^{*}=k_{\rm B}T/\epsilon$, $k_{\rm B}$ 
being the Boltzmann constant, it follows that the only parameter affecting 
the phase diagram is the ratio of the interaction strengths between unlike and 
like species $\delta=\epsilon_{12}/\epsilon$. 

Clearly, symmetric mixtures appear
quite artificial when considered as a model of real binary fluids. In fact,
some of the features of their phase behavior hinge on the invariance with
respect to the exchange of the two species, and are not found in the phase
diagram of real mixtures. As has already been observed~\cite{wilding}, 
symmetric mixtures are better seen as a model for a one-component fluid,
whose particles have been endowed with a two-state, spin-like variable 
in addition to their translational degrees of freedom, so that their mutual
interaction depends both on their relative position and on their ``internal''
state, namely whether the interacting particles belong to the same species or 
not. As such, this model mixture is closely related to other models of 
dipolar~\cite{zhang,groh} and magnetic~\cite{hemmer,tavares,oukouiss,zaluska} 
fluids, especially Ising-spin fluids~\cite{nijmeijer,schinagl,sokolovskii}. 
In these systems, the phase behavior results from
the interplay between the liquid-vapor phase separation and the additional 
transition, e.g. para-ferromagnetic, associated to the spin-like degrees 
of freedom. In symmetric mixtures, the latter corresponds to the 
mixing-demixing transition. 

Because of the relative simplicity of this model
compared to a generic binary mixture and of the possibility of generating  
the whole spectrum of phase diagrams by acting on just one parameter,
symmetric mixtures have been widely studied both by mean-field 
theory~\cite{wilding} and by numerical 
simulations~\cite{wilding,panagiotopulos,recht,green,gama,wilding2,costa}. 
A situation which has been
given special attention is that of equal species concentration 
$x=\rho_{2}/(\rho_{1}+\rho_{2})=1/2$. In this case, accurate numerical 
simulations~\cite{wilding} have qualitatively confirmed the mean-field 
scenario for 
the phase diagram as the parameter $\delta$ is varied in the interval 
$0<\delta<1$. This will be considered in  
detail in Secs.~\ref{sec:results1}, \ref{sec:results2}. Here we just recall 
that the mean-field phase diagram at equal
species concentration presents both a first-order coexistence 
boundary which separates a low-density fluid from a high-density one, and 
a line (the so-called $\lambda$-line) of mixing-demixing critical points. 
Beyond the $\lambda$-line, 
the fluid actually consists of two demixed fluids
in equal amounts, one at a certain concentration $\overline{x}$, and the other
at concentration $1-\overline{x}$, so that the overall concentration of the two 
species remains the same. 
For large enough $\delta$ ($\delta_{1}<\delta<1$ with $\delta_{1}=0.708$
according to the mean-field result~\cite{wilding})
the coexistence curve ends into a liquid-vapor critical point, while the
$\lambda$-line intersects the coexistence curve at a point of first-order 
transition, thereby terminating into a critical endpoint. 
At small $\delta$ ($0<\delta<\delta_{2}$ with 
$\delta_{2}=0.605$ in mean field~\cite{wilding}),
on the other hand, the point at which the coexistence curve meets the 
$\lambda$-line coincides with its critical point. The latter is then 
referred to as a tricritical point, since on approaching this point from 
low temperatures, one observes the simultaneous coalescence of three phases, 
namely the low-density vapor and the two demixed high-density fluids. 
Finally, in a narrow interval of $\delta$ values 
$\delta_{2}<\delta<\delta_{1}$ intermediate between those corresponding to 
the two topologies described above, one has the occurrence of both 
a liquid-vapor critical point and a tricritical point. 
Despite the qualitative agreement between the mean-field scenario and 
the simulation results, there are still several points that deserve further 
investigation. First, mean-field theory and simulations show considerable
quantitative discrepancies, which concern both the position of the critical 
loci and the values of $\delta$ at which the changes in the topology of 
the phase diagram occur. This is in itself not surprising, as mean-field
theory cannot be expected to be quantitatively very accurate. Therefore, 
one would like 
to go beyond it by means of a theoretical treatment which includes 
fluctuations in the order parameter of the transition, be it of 
the liquid-vapor or mixing-demixing kind. Two relevant issues in this respect
are whether the mean-field scenario is qualitatively recovered even after
fluctuations have been taken into account, and which is the extent of the
quantitative changes involved. Moreover, the case of equal species 
concentration corresponds just to a certain plane, albeit undoubtedly 
of special interest, of the space of thermodynamic states. Simulation
studies of the phase diagram have indeed been performed also at fixed density
$\rho$ and variable concentration $x$~\cite{gama,costa}, but mapping 
the phase diagram in the
whole thermodynamic space for different values of the parameter $\delta$ 
would require an exhorbitant number of simulation runs, and therefore hardly
appears as a viable strategy in view of the computer time required. It is then 
tempting to resort to theory in order to explore the phase diagram at general
density and concentration and find out how it changes by changing $\delta$,
so as to see what the phase diagrams corresponding to the three regimes 
outlined above look like, as one moves away from the $x=1/2$ plane.  

We aim to address these topics in the present work by means of the 
hierarchical reference theory (HRT) of binary fluids. This theory has already
been successfully applied to the description of the critical 
behavior~\cite{hrt1}, crossover phenomena~\cite{hrt1,hrt2}, and phase 
diagram~\cite{hrt3} of simple fluid mixtures. Few results for the phase 
behavior of symmetric mixtures have also been reported~\cite{hrt3,const}, but 
a systematic investigation of this system by HRT has not been undertaken yet. 
In our opinion, HRT is especially well suited for such a study, with 
particular regard for the tasks stated above. In fact, the aim of this 
approach is determining how the Helmholtz free energy of the mixture 
is affected by the introduction of density and concentration fluctuations. 
This is achieved via a renormalization-group (RG) like procedure, where the 
long-wavelength Fourier components of the microscopic interaction are 
gradually introduced into the hamiltonian of the mixture. Any intermediate
stage of this process, such that only Fourier components with wavevectors 
exceeding a certain cutoff $Q$ have been taken into account, physically 
corresponds to suppressing fluctuations on a lengthscale $L>1/Q$. Long-range
fluctuations are recovered in the limit $Q\rightarrow 0$, when the free energy
of the fully interacting system is obtained, while the mean-field free energy
enters as the initial condition at $Q=\infty$. The main advantage of HRT over 
other liquid-state theories is that it embodies several features of the 
RG description of critical phenomena in a treatment based
on the microscopic hamiltonian of the fluid. These include scaling, 
non-trivial critical exponents, and the correspondence between universality
classes and different fixed points of the RG flow~\cite{hrt1}. 
A fact of particular relevance for the investigation pursued here is that 
the inclusion of long-range fluctuations has the effect of preserving 
the correct convexity of the free energy in the whole thermodynamic space.
Whenever phase coexistence occurs, one does not find any domain of instability 
as in the mean-field approximation, and the conditions of thermodynamic 
equilibrium between the phases at coexistence are enforced by the theory 
itself. At each given temperature, the coexistence region is then immediately
recovered as the locus in the density-concentration plane where the chemical 
potential of each component is constant along the lines of fixed pressure, 
with no need of imposing this condition {\em a posteriori} by a Maxwell 
construction. For binary mixtures, the latter proves to be quite cumbersome
already at the mean-field level, and is much more so for more sophisticated 
integral-equation theories, in which the occurrence of phase separation 
generally entails the presence of some forbidden domain, where the theory 
cannot be solved at all. Therefore, the ability of straightforwardly mapping
the phase diagram is a valuable asset of HRT. This is especially true in the 
present case where the topology of the phase diagram is very sensitive to
changes in $\delta$, while at the same time, as will be seen in the following, 
the features that allow one to discriminate between different topologies are
often detectable only in a narrow window of the thermodynamic space. Because
of the lack of a solution defined for every $(T, \rho, x)$ state, 
pinpointing all these features by conventional 
integral-equation theories would undoubtedly 
prove extremely difficult, perhaps even impossible. 

In this work we have considered symmetric mixtures of additive hard spheres
interacting via an attractive Yukawa tail potential 
$w_{ij}(r)=-\sigma\epsilon_{ij}e^{-z(r/\sigma-1)}/r$, where $r$ is the 
interparticle distance and $z$ is the inverse range of the interaction, which 
has been fixed to the value $z=1.8$ for both like and unlike species. This 
hard-core Yukawa (HCY) form has been preferred to the square-well potential 
used in 
Ref.~\cite{wilding}. The latter lends itself well to simulation, but the very
slowly decaying behavior of its Fourier transform makes it somewhat tedious 
to use in HRT. 
The HCY potential has already been adopted in a number of studies 
of symmetric mixtures based on the mean spherical approximation 
(MSA)~\cite{kahl1}, the optimized random phase approximation
(ORPA)~\cite{kahl2}, and the self-consistent Ornstein-Zernike 
approximation (SCOZA)~\cite{kahl3}, all of which yielded 
for the phase diagram at equimolar concentration 
the same behavior found in mean-field theory. 
The scenario that comes out of our investigation by HRT agrees 
qualitatively with the mean-field one
in predicting that, as $\delta$ is increased, the phase diagram at equimolar
concentration exhibits first a tricritical point, and subsequentely 
both a tricritical and a liquid-vapor critical point. However, we did 
not find any clear evidence of the occurrence of the mixing-demixing critical 
endpoint given by mean-field theory as $\delta$ approaches $1$,
at least up to $\delta=0.8$, above which further investigation is hindered 
by the finite resolution of the density grid used 
in our numerical calculation.
Besides the scenario sketched above at equal species
concentration, 
other interesting 
features of the phase diagram emerge as we move off the $x=1/2$ axis. 
For instance, the intermediate regime where both a critical and a tricritical 
point are found for $x=1/2$ is marked in the $\rho$-$x$ plane by a coexistence
region which in a certain temperature interval contains a ``hole'' or 
``island'' of homogeneous, mixed fluid. Double critical 
points and, for a certain value of $\delta$, tricritical points are also 
observed for unequal species concentration. 

The paper is structured as follows: the HRT for a binary fluid is described
in Sec.~\ref{sec:theory}. Mean-field theory is recovered within this approach 
as a zeroth-order approximation,  
and the mean-field results for the critical lines of a symmetric mixture
as the parameter $\delta$ is varied in the interval $0<\delta<1$ are shown 
in Sec.~\ref{sec:results1}. The HRT phase diagram for different values of 
$\delta$ is discussed and compared with the mean-field predictions in 
Sec.~\ref{sec:results2}. Finally, 
in Sec.~\ref{sec:conclusions} we summarize our findings and draw 
our conclusions.           

\section{HRT equations}
\label{sec:theory}

Here we briefly review the HRT approach for a binary fluid. A more detailed
derivation can be found in previous works~\cite{hrt1,hrt3,hrt4}.

We consider a model mixture consisting of particles of two species interacting 
via a two-body spherically symmetric potential $v_{ij}(r)$, where the indices
$i$, $j$ label the particle species. The derivations presented in this
Section do not hinge on the fact that one is dealing with a symmetric system
such that $v_{11}(r)=v_{22}(r)$, so they are equally valid for non-symmetric 
systems. We assume that $v_{ij}(r)$ can be split as the sum of a singular
contribution $v^{\rm R}_{ij}(r)$ which accounts for the short-range repulsion 
between the particles, and a longer-ranged, smooth
attractive tail $w_{ij}(r)$ which may induce fluid-fluid phase separation. 
The properties of the mixture interacting via the repulsive potential 
$v^{\rm R}_{ij}(r)$ alone are considered as known, so that it acts as 
a ``reference'' or unperturbed system. For a hard-core plus tail potential, 
the natural choice for the 
reference system is just a binary mixture of additive hard spheres, which can
be described by the Mansoori-Carnahan-Starling-Leland equation 
of state~\cite{mansoori} and the corresponding Verlet-Weis parameterization 
for the two-body correlations~\cite{grundke,lee}. We remark that here we will 
not 
be concerned with a possible demixing transition occurring in the hard-sphere
mixture, since such a transition may come along only as a consequence of 
depletion interactions when the particles differ widely in size. 
In the present case of equisized particles, the reference system 
reduces to a one-component hard-sphere fluid, 
and all the fluid-fluid transitions
displayed by the system are necessarily due to the attractive perturbation
$w_{ij}(r)$. The HRT differs from the conventional liquid-state approaches 
in the way this perturbation is dealt with. In order to accurately describe
the long-range fluctuations that are important in criticality and phase 
separation, the attractive part of the interaction is switched on gradually
by introducing a $Q$-{\em system} with a modified interaction 
$v^{Q}_{ij}(r)=v^{\rm R}_{ij}(r)+w^{Q}_{ij}(r)$, where $w^{Q}_{ij}(r)$ 
is defined in such a way that its Fourier transform 
$\widetilde{w}^{Q}_{ij}(k)$ coincides with that of the original attractive 
potential $\widetilde{w}_{ij}(k)$ for $k>Q$, and vanishes for $k<Q$. 
Inspection of the diagrammatic series of the Helmholtz free energy 
of the mixture in terms of the perturbation $w_{ij}(r)$ and the correlation 
functions of the reference system shows that introducing such an infra-red 
cutoff in the interaction is physically equivalent to inhibiting fluctuations
with characteristic lengths $L>1/Q$. If $Q$ is made evolve from $Q=\infty$,
the $Q$-systems evolve from the reference system by acquiring fluctuations
of longer and longer wavelengths. The fully interacting system is recovered
as the $Q\rightarrow 0$ limit of such a process. Only in this limit true 
long-range correlations are allowed to develop in the fluid. The equation 
for the corresponding
evolution of the Helmholtz free energy $A_{Q}$ of the $Q$-systems can be 
determined exactly and is related to the attractive perturbation in momentum
space $\Phi_{ij}(k)=-\beta\widetilde{w}_{ij}(k)$ where $\beta=1/(k_{\rm B}T)$,
and to the direct correlation function of the $Q$-system in momentum space 
$c^{Q}_{ij}(k)$. We recall that the direct correlation function is related
to the structure factor of the fluid by the Ornstein-Zernike relation. 
If we denote by $c_{Q}(k)$ the $2\times 2$ symmetric matrix with elements 
$c^{Q}_{ij}(k)$, for a binary mixture this relation reads
\begin{equation}
\left[c^{-1}_{Q}(k)\right]_{ij}=
-\sqrt{\rho_{i}\rho_{j}}\, S^{Q}_{ij}(k) \, ,
\label{oz}
\end{equation}    
where $c^{-1}_{Q}(k)$ is the inverse matrix of $c_{Q}(k)$ and $S^{Q}_{ij}(k)$
is the partial structure factor of the $Q$-system. Here, unlike in the
convention commonly adopted in liquid-state theory, $c^{Q}_{ij}(k)$ contains
its ideal-gas contribution $-\delta_{ij}/\rho_{i}$. The evolution equation
for the Helmholtz free energy is most easily formulated in terms of a modified 
free energy ${\scr A}_{Q}$ and direct correlation function 
${\scr C}_{Q}(k)$
defined as:
\begin{eqnarray}
{\scr A}_{Q} & = & -\frac{\beta A_{Q}}{V}+\frac{1}{2}\sum_{i,j=1}^{2}
\rho_{i}\rho_{j}\left[\Phi_{ij}(k\!=\!0)-\Phi^{Q}_{ij}(k\!=\!0)\right]
-\frac{1}{2}
\sum_{i=1}^{2}\rho_{i}\int\!\!\frac{d^{3}{\bf k}}{(2\pi)^{3}}
\left[\Phi_{ii}(k)-\Phi^{Q}_{ii}(k)\right]  
\label{amod} \\
{\scr C}^{Q}_{ij}(k) & = & c^{Q}_{ij}(k)+\Phi_{ij}(k)-\Phi^{Q}_{ij}(k) \, .
\label{cmod}
\end{eqnarray}    
The HRT equation for ${\scr A}_{Q}$ is then
\begin{equation}
\frac{\partial {\scr A}_{Q}}{\partial Q}=-\frac{Q^2}{4\pi^2}
\log\left\{\det\left[{\bf 1}-
{\scr C}^{-1}_{Q}(Q)\Phi(Q)\right]\right\} \, ,
\label{hrt}
\end{equation}   
where again $\Phi(Q)$ and ${\scr C}^{-1}_{Q}(Q)$ are $2\times 2$ symmetric 
matrices, the latter being the inverse of the matrix ${\scr C}_{Q}(Q)$. 
We note that for $Q\rightarrow 0$, i.e. at the end of the evolution process,
$\Phi_{Q}(k)$ and $\Phi(k)$ coincide, so that in this limit the modified 
quantities ${\scr A}_{Q}$, ${\scr C}_{Q}$ yield respectively the true free 
energy and direct correlation function of the fully interacting system.   
For $Q=\infty$ instead one has $\Phi_{Q}(k)\equiv 0$, and ${\scr A}_{Q}$ 
and ${\scr C}_{Q}$ are nothing but the mean-field free energy and the 
random-phase approximation (RPA) direct correlation function in the presence 
of the full perturbing potential $\Phi(k)$. These play the role of the initial
conditions of the evolution equation~(\ref{hrt}), which then
describes how 
the mean-field estimate for the free energy is affected by the inclusion 
of fluctuations. This equation is manifestly not closed, since the evolution
of the free energy ${\scr A}_{Q}$ is related to the matrix of the direct 
correlation function ${\scr C}_{Q}(k)$, which is itself unknown. 
In fact, Eq.~(\ref{hrt}) is just the first equation of an infinite hierarchy
for the direct correlation functions of increasing order: for instance, 
the evolution of ${\scr C}_{Q}(k)$ is related to the $3$- and $4$-body direct
correlation functions in Fourier space~\cite{hrt4}. A point of crucial 
importance in order
to implement a viable HRT scheme consists then in supplementing 
Eq.~(\ref{hrt}) with some closure relation involving ${\scr C}_{Q}(k)$. 
Here, as well as in the previous applications of HRT, we have not resorted to
the higher-order equations of the hierarchy. Instead, we have adopted for
${\scr C}_{Q}(k)$ an approximate form inspired by standard perturbative
liquid-state theories:
\begin{equation}
{\scr C}^{Q}_{ij}(k)=c^{\rm HS}_{ij}(k)+\nu^{Q}_{ij}\,\Phi_{ij}(k) \, ,
\label{closure}
\end{equation}
where $c^{\rm HS}_{ij}(k)$ is the Fourier transform of the partial direct
correlation function of the hard-sphere reference system, which has been 
represented by the above-mentioned Verlet-Weis parameterization. 
The functional form of Eq.~(\ref{closure}) for the direct correlation function
is similar to that of the widely used RPA, which is recovered 
for $\nu^{Q}_{ij}=1$. In particular, both of them rely
on the Ornstein-Zernike {\em ansatz}, i.e., the direct correlation function 
has always the same range as the potential, so that ${\scr C}^{Q}_{ij}(k)$
is always analytic in $k$, including at the critical points of the system,
where the real direct correlation function is instead expected to be 
non-analytic for $Q\rightarrow 0$. However, unlike in the RPA, the amplitude
$\nu^{Q}_{ij}$ of the perturbation is regarded as an unknown quantity, to be
determined in such a way that each $Q$-system satisfies the compressibility
sum rule. In a one-component fluid, this relates the isothermal 
compressibility to the zero-wavevector value of the structure factor, or 
equivalently of the direct correlation function in momentum space. Such a rule
is readily generalized to binary systems, and in terms of the modified 
quantities ${\scr A}_{Q}$, ${\scr C}_{Q}(k)$ it reads
\begin{equation}
{\scr C}^{Q}_{ij}(k\!=\!0)=\frac{\partial^{2}{\scr A}_{Q}}{\partial \rho_{i}
\partial \rho_{j}} \mbox{\hspace{2cm}} i,j=1,2 \, .
\label{sum}
\end{equation}   
By determining $\nu^{Q}_{ij}$ so that Eq.~(\ref{sum}) is satisfied, one 
immediately finds from Eq.~(\ref{closure})
\begin{equation}
{\scr C}^{Q}_{ij}(k)=c^{\rm HS}_{ij}(k)+\left[\frac{\partial^{2}{\scr A}_{Q}}
{\partial \rho_{i} \partial \rho_{j}}-c^{\rm HS}_{ij}(k\!=\!0)\right]
\varphi_{ij}(k)
\, ,
\label{closure2}  
\end{equation}
where we have set $\varphi_{ij}(k)=\Phi_{ij}(k)/\Phi_{ij}(k\!=\!0)$. 
Eq.~(\ref{closure2}) ensures that the Helmholtz free energy obtained by 
integration of Eq.~(\ref{hrt}) is consistent with the compressibility route
to the thermodynamics given by Eq.~(\ref{sum}). This thermodynamic consistency
condition plays a key role in the implementation of HRT. In fact, 
by substituting ${\scr C}^{Q}_{ij}(k)$ as given by Eq.~(\ref{closure2})
into the HRT equation~(\ref{hrt}), a closed partial differential equation 
(PDE) for ${\scr A}_{Q}$ is obtained, which involves both the first partial
derivative of ${\scr A}_{Q}$ with respect to $Q$, and its second partial 
derivatives with respect to the densities $\rho_{1}$, $\rho_{2}$. In order
to integrate this equation numerically, we found it most convenient to cast
it into a form where the partial derivatives of the unknown function appear
only outside some ``coefficients'' that may depend both on the independent 
variables $\rho_{1}$, $\rho_{2}$, $Q$, and on the unknown function itself,
but do not contain its derivatives. This allows us to take advantage of
finite-difference schemes especially devised for equations of such a 
{\em quasi-linear} form, which combine robustness with a moderate 
computational cost~\cite{ames}. Both of these are very important requirements 
in our case.
In fact, in order to deal with the divergence of the compressibility 
at criticality and phase coexistence, one has to resort to a very stable 
algorithm, while on the other hand for a diffusive PDE in three independent
variables like the one considered here, the dimension of the vectors 
generated by the discretization procedure becomes rapidly very large even for 
a relatively coarse density step $\Delta\rho$. Hence, we have to use 
a solution scheme which is not too demanding computationally to prevent 
computer time from increasing beyond control. 

The method we adopted to re-write Eq.~(\ref{hrt}) supplemented by the closure
relation~(\ref{closure2}) in quasi-linear form has already been illustrated
in Ref.~\cite{hrt3}, to which we refer the reader for details. 
Here we recall that, instead of ${\scr A}_{Q}$, we use as unknown function
the quantity
\begin{equation}
U=\log\left\{\det\left[{\bf 1}-
{\scr C}^{-1}_{Q}(Q)\Phi(Q)\right]\right\} \, .
\label{u}
\end{equation}  
This is found to satisfy a quasi-linear PDE of the form
\begin{equation}
e^{U}\,\frac{\partial U}{\partial Q}=K\frac{\partial^{2}U}
{\partial\rho_{1}^{2}}+
L\frac{\partial^{2}U}{\partial\rho_{1}\partial\rho_{2}}+
M\frac{\partial^{2}U}{\partial\rho_{2}^{2}}+N 
\label{pde}
\end{equation}  
whose ``coefficients'' $K$, $L$, $M$, $N$, which will not be reported here,
depend on the variables $\rho_{1}$, $\rho_{2}$, $Q$ both 
explicitly and implicitly via a set of three auxiliary variables. 
These can be identified either with the eigenvalues $\lambda_{1}$, 
$\lambda_{2}$ of the symmetric matrix ${\scr C}_{Q}(Q)$ and the angle 
$\alpha$ of the rotation that casts ${\scr C}_{Q}(Q)$ into diagonal form,
or with the corresponding quantities $\Lambda_{1}$, $\Lambda_{2}$, 
$\theta$ for the symmetric matrix of elements  
$\partial^{2}{\scr A}_{Q}/\partial\rho_{i}\partial\rho_{j}$. 
As discussed in~\cite{hrt3}, the use of one or the other set of auxilary 
variables is dictated by the different behavior of the interaction in the
high- and low-$Q$ region. 
At high $Q$, where 
the Fourier transform of the attractive perturbation $\varphi_{ij}(Q)$
typically displays oscillations, thereby vanishing for certain $Q$ values,
it is better to adopt as auxiliary 
quantities $\Lambda_{1}$, $\Lambda_{2}$, $\theta$. At small $Q$ instead,
when most of the attractive interaction has been included into the system
and phase coexistence may occur, the set $\lambda_{1}$, $\lambda_{2}$, 
$\alpha$ is the more appropriate choice. In both cases, the derivative 
of each of the auxilary variables with respect to $Q$ can be expressed 
in terms of the variables themselves and the partial derivatives of $U$ 
with respect to $\rho_{1}$, $\rho_{2}$. At any given $Q$, the resulting
equations are used to update $\lambda_{1}$, $\lambda_{2}$, $\alpha$ 
(or $\Lambda_{1}$, $\Lambda_{2}$, $\theta$), while the PDE~(\ref{pde})
is used to update $U$. Since the matrix ${\scr C}_{Q}(Q)$ that appears 
in Eq.~(\ref{u}) can be expressed either by $\lambda_{1}$, $\lambda_{2}$, 
$\alpha$ or, via Eq.~(\ref{closure2}), $\Lambda_{1}$, $\Lambda_{2}$, $\theta$,
it follows that $U$ and the set of the three auxiliary variables are not
mutually independent. The relation between $U$ and the auxiliary variables
is used throughout the integration procedure as a check of the accuracy 
of the calculation.      

In order to perform the numerical integration, we found it convenient to
replace the independent variables $\rho_{1}$, $\rho_{2}$ with the related 
variables $\xi=(\rho_{1}+\rho_{2})\sigma^{3}$, 
$x=\rho_{1}/(\rho_{1}+\rho_{2})$. At any given $Q$, $U$ is then defined 
in the rectangular domain $0\leq x\leq 1$, $0\leq \xi\leq\xi_{\rm max}$, where
the high-density boundary $\xi_{\rm max}$ was chosen equal to $1$. 
At the beginning of the evolution process, i.e. for suitably large $Q$, 
the matrix
${\scr C}_{Q}(k)$ can be identified with the RPA expression for the direct
correlation function which, as said above, is obtained from Eq.~(\ref{cmod})
for $Q=\infty$. This gives the initial condition for $U$ via 
Eq.~(\ref{u}). The boundary conditions for $U$ needed for the integration of
Eq.~(\ref{pde}) are determined as follows: for $\xi=0$ the diagonal elements 
of the matrix ${\scr C}_{Q}(Q)$ diverge because of the ideal-gas contribution
to the direct correlation function $-\delta_{ij}/\rho_{i}$ , so that 
$U(\xi=0,x)$ vanishes identically. For $x=0$ and $x=1$ only one of the two
species is present. The corresponding boundary conditions are then given by 
the solution of the HRT equation for a pure fluid, which can be integrated
numerically by specializing the procedure sketched above to a one-component
system.
The high-density boundary condition for $\xi=\xi_{\rm max}$ is non-trivial
because, unlike in pure fluids, we must allow the occurrence of phase 
transitions even at high density. 
As a consequence, we have to rule out the possibility of using 
for $U$ at $\xi=\xi_{\rm max}$ a simple approximation such as the RPA, as 
already done for the one-component case, because such a form would behave 
unphysically in the coexistence region. In general, we expect that at high 
density the compressibility of the fluid will be small, and concentration
fluctuations will become more important than density ones. This corresponds 
to the transition becoming less liquid-vapor and more mixing-demixing 
in character. For the symmetric mixtures considered here, such an expectation
holds rigorously: in fact, because of the special symmetry of the model,
the high-density transition are of pure mixing-demixing type. 
As a consequence, 
we expect $U$ to be much more sensitive
to a change in concentration than in density. Accordingly, 
for $\xi=\xi_{\rm max}$ the partial derivatives of $U$ along 
$\xi$ in Eq.~(\ref{pde}) have been disregarded with respect to those along 
the conjugate direction $z=(\rho_{1}-\rho_{2})\sigma^{3}$.
By switching from the variables $\xi$, $z$ back to the variables 
$\xi$, $x$, we obtain a PDE in the variables $Q$, $x$ 
on the boundary $\xi=\xi_{\rm max}$, 
whose solution yields the high-density boundary condition of Eq.~(\ref{pde}). 

The numerical integration of Eq.~(\ref{pde}) supplemented by the equations
for the evolution of the auxilary variables was performed 
on a grid in the $(\xi, x)$
plane containing $150\times 150$ mesh points. The integration 
with respect to the variable $Q$ was carried out by setting 
$Q=Q_{0} e^{-t}$, $t\geq 0$, where the initial value of $Q$ corresponding 
to $t=0$ was typically fixed at $Q_{0}=30\, \sigma^{-1}$. The variable $t$ was
discretized using a step $\Delta t=10^{-3}$ and the iteration in $t$ went on 
until convergence in the quantity $U$ was achieved outside the coexistence
region. At low temperature this requirement can be satisfied for 
$Q<10^{-4}\sigma^{-1}$.   

\section{Mean-field critical lines}
\label{sec:results1}

The phase diagrams of binary mixtures are usually classified 
according to the topology of their critical lines\cite{konynenberg,rowlinson}. 
We then 
begin the discussion of the phase behavior of the symmetric mixtures as 
a function of the interaction parameter $\delta$ by presenting the different 
shapes of the critical lines that are predicted by the mean-field 
approximation.  
We will then consider the HRT results for the phase diagram and elucidate 
the relationship between the critical loci and the behavior of the coexistence
domains on changing the temperature. In doing so, we will also be 
in a position to compare the mean-field and HRT predictions. 

As observed above, the mean-field (MF) approximation enters the HRT as
the initial condition at $Q=\infty$, when no fluctuations have been introduced
into the system:
\begin{equation}
-\frac{\beta A_{\rm MF}}{V}={\scr A}_{Q=\infty}=
-\frac{\beta A_{\rm HS}}{V}+\frac{1}{2}\sum_{i,j=1}^{2}
\rho_{i}\rho_{j}\Phi_{ij}(k\!=\!0)
-\frac{1}{2}
\sum_{i=1}^{2}\rho_{i}\int\!\!\frac{d^{3}{\bf k}}{(2\pi)^{3}}\Phi_{ii}(k) \, .
\label{ameanfield}
\end{equation}
The equations for the MF critical loci are also obtained within HRT as
the lowest-order approximation to the requirement that the RG flow generated
by the evolution equation~(\ref{hrt}) drives the free energy of the mixture
towards its fixed point. This gives the following equations~\cite{hrt1}:
\begin{eqnarray}
& &\frac{\partial^{2}}{\partial\psi_{1}^{2}}
\left(\frac{\beta A_{\rm MF}}{V}\right) = 0 \, ,
\label{mf2} \\
& &\frac{\partial^{3}}{\partial\psi_{1}^{3}}
\left(\frac{\beta A_{\rm MF}}{V}\right) = 0 \, ,
\label{mf3} \\
& &\frac{\partial^{2}}{\partial\psi_{1}\partial\psi_{2}}       
\left(\frac{\beta A_{\rm MF}}{V}\right) = 0 \, ,
\label{mf1} 
\end{eqnarray}
where $\psi_{1}$, $\psi_{2}$ are obtained from the original densities
$\rho_{1}$, $\rho_{2}$ via an {\em a priori} unknown rotation such that 
Eq.~(\ref{mf1}) is satisfied. 
Eqs.~(\ref{mf2}), (\ref{mf3}) are formally similar to
the equations for the critical point of a pure fluid, except that here
the density $\rho$ has been replaced by the variable $\psi_{1}$. We note that
Eqs.~(\ref{mf2}), (\ref{mf1}) amount to requiring that the hessian determinant 
of the free energy must vanish at the critical point, and that the vanishing
eigenvalue corresponds to the eigenvector directed along $\psi_{1}$. 
This is the linear combination of the densities that gives the direction 
of strongest fluctuation, and identifies the order parameter 
of the transition. For the solutions of Eqs.~(\ref{mf2})--(\ref{mf1}) 
to yield actual critical points,
a further constraint has to be imposed, i.e. the condition of thermodynamic
stability ensuring that 
even a critical point is an equilibrium state of the system, 
and as such is stable against density and concentration fluctuations. 
The stability conditions at a critical point read~\cite{hrt1}
\begin{eqnarray}
& &\frac{\partial^{2}}{\partial\psi_{2}^{2}}
\left(\frac{\beta A_{\rm MF}}{V}\right) > 0 \, ,
\label{mf4} \\
& &\left[\frac{\partial^{3}}{\partial\psi_{1}^{2}\partial\psi_{2}}
\left(\frac{\beta A_{\rm MF}}{V}\right)\right]^{2}-\frac{1}{3}
\frac{\partial^{2}}{\partial\psi_{2}^{2}}
\left(\frac{\beta A_{\rm MF}}{V}\right)
\frac{\partial^{4}}{\partial\psi_{1}^{4}}
\left(\frac{\beta A_{\rm MF}}{V}\right) < 0 \, .
\label{mf5}   
\end{eqnarray}   
The three equations~(\ref{mf2})--(\ref{mf1}) with the conditions (\ref{mf4}),
(\ref{mf5}) contain four unknowns, namely $\rho_{1}$, 
$\rho_{2}$, $T$, and the state-dependent angle $\gamma$ of the rotation 
that identifies the proper axes $\psi_{1}$, $\psi_{2}$. 
As $\gamma$ is varied, they will generate a set of lines in the thermodynamic 
space. We must observe that, when several critical lines are present,
one should also check that a critical point does not fall into the coexistence
region originating from another critical line. In mean-field theory,
this circumstance can occur without violating the stability conditions 
(\ref{mf4}), (\ref{mf5}), when the point considered lies between the binodal
and the spinodal surfaces of a neighboring transition. In such a situation,
a solution of Eqs.~(\ref{mf2})--(\ref{mf5}) corresponds to a critical point
which, while still locally stable, is however globally metastable with 
respect to first-order phase separation. In order to assess this possibility, 
one should then determine the mean-field binodal surfaces by a Maxwell 
construction. This has not been done here. However, we can discriminate
between stable and metastable regimes, at least for equimolar concentrations,
by comparing our results with those obtained in Ref.~\cite{wilding}, where 
mean-field binodals were determined. We recall that at the mean-field level
the phase behavior as a function of $\delta$ is independent of the profile 
of the attractive interaction $\Phi_{ij}$, since this enters 
in the approximation only via its spatial integral. A change in the form 
of $\Phi_{ij}$ is then taken into account by simply rescaling the temperature. 

Let us now consider the critical lines of the symmetric mixtures we are 
interested in. Because of the attractive interaction between 
the particles, we expect that for suitable temperature, density, 
and concentration, the system will undergo a liquid-vapor transition. 
This is certainly true for states at low or high
concentration, where one of the two species will play a minor role. 
On the other hand, since the interaction between unlike particles is weaker
than that between like particles, the internal energy will tend to promote
demixing between the two species. For high enough density, this increase 
in the absolute value of the internal energy may overcome the loss in entropy
implied by the demixing, and a mixing-demixing transition may appear. 
Therefore, both liquid-vapor and mixing-demixing critical lines are expected.
As the unlike-to-like interaction ratio $\delta$ is lowered, 
the mixing-demixing transition becomes more favored, and the corresponding 
critical line will move to lower density. The projections of the mean-field
critical lines on the density-concentration plane for a relatively small 
value of $\delta=0.4$ are reported in Fig.~\ref{fig:mf1}. The open dots 
mark the positions of the minima of the critical temperature, while 
the arrows drawn
along the critical lines give a graphic representation of the relative weight
of density and concentration fluctuations, thereby showing the direction 
of the order parameter $\psi_{1}$ determined by Eqs.~(\ref{mf2})--(\ref{mf1}).
Specifically, the angle $\phi$ between the arrows and the density axis gives 
the fluctuation of the order parameter corresponding to a given fluctuation
of the total density $\rho$ and concentration $x$ as 
$\delta \psi_{1}=\delta \rho \cos\!\phi+\rho\, \delta x \sin\!\phi$. 
Arrows parallel to the $\rho$- and $x$-axis then indicate pure liquid-vapor
and mixing-demixing transitions, respectively. Fig.~\ref{fig:mf1} shows
a mixing-demixing critical line at equal species concentration $x=1/2$. 
On the low-density side, this intersects another critical line that connects
the critical points of the pure species. As one moves from the pure species
to the equimolar mixture, the direction of the order parameter changes
continuously from pure liquid-vapor to pure mixing-demixing. The critical 
temperature, not shown in the figure (see the lower panel of 
Fig.~\ref{fig:crit} for the similar case $\delta=0.5$), initially decreases 
until it reaches 
a minimum at two points symmetric with respect to $x=1/2$, after which 
it starts increasing and keeps on doing so along the mixing-demixing line
as the density is increased. This behavior of the critical temperature 
is related to the change in the order parameter of the transition. Close 
to $x=0$ or $x=1$, where phase separation is essentially of liquid-vapor type,
increasing the amount of the dilute component increases the weight of the 
interaction between unlike species in the internal energy. Since here we have
$\delta<1$, this leads to a decrease of the overall attractive contribution
to the internal energy, resulting in a lower critical temperature 
of the liquid-vapor transition. On the other hand, the same argument implies 
that for a transition which is mainly mixing-demixing in character, 
approaching equimolar concentration increases the gain in internal energy
entailed by the demixing, so that the critical temperature increases as one
moves towards $x=1/2$. Once the mixing-demixing critical line has been 
reached, an increase of the density at constant concentration similarly favors
the energetic contribution to the free energy and leads to an increase 
of the critical temperature. 

As $\delta$ is increased, the mixing-demixing line, as noted above, moves to 
higher density, and so does that portion of the line originating from the pure
species where the transition is predominantly of demixing type. At about
$\delta=0.46$, a new feature appears in the critical lines, namely 
a crescent-shaped line at low density and concentration spanning an interval
centered at $x=1/2$, where the transition is essentially liquid-vapor. 
This is shown in Fig.~\ref{fig:mf2} for $\delta=0.65$. The occurrence 
of such a critical line can be euristically explained as follows: if $\delta$
were equal to $1$, the mixture would reduce to a one-component hard-sphere 
fluid with attractive tail interaction. If $\delta$ is not too small, a nearly
equimolar mixture may still behave like a sort of ``effective'' one-component
fluid displaying a liquid-vapor transition. In order for this to happen,
however, the density has to be low enough, so that the smaller gain 
in internal energy resulting from choosing a liquid-vapor phase separation 
at intermediate concentration instead of a mixing-demixing one can be 
compensated by a larger entropy.
We must point out that, on the basis of the investigation performed 
in Ref.~\cite{wilding} and the discussion made here below 
Eqs.~(\ref{mf2})--(\ref{mf5}), we do not expect this critical line to appear 
in the equilibrium 
phase diagram right above $\delta=0.46$. In fact, the interval 
$0.46< \delta < 0.605$ corresponds to the hidden-binodal regime 
of Ref.~\cite{wilding}, where the critical point at $x=1/2$ 
of the crescent-shaped line is metastable. Of course, knowledge 
of the behavior of the point at equimolar concentration alone 
is not sufficient to deliberate about the fate of the whole critical line. 
In principle, some portions of it might become stable for $\delta$ different 
from the value $\delta=0.605$ reported in~\cite{wilding}. However, this would 
imply a change in the topology of the critical lines with respect to that
shown in Fig.~\ref{fig:mf2} which we did not observe when fluctuations are 
taken into account (see Sec.~\ref{sec:results2}), and we regard such 
an occurrence as rather unlikely. In summary, it is palusible that the value 
$\delta=0.605$ obtained in Ref.~\cite{wilding} sets the threshold 
for the appearance in the mean-field equilibrium phase diagram of the whole 
crescent-shaped critical line besides the point at $x=1/2$.       
The critical temperature along 
the crescent-shaped line changes very little, typically by few percent, 
and it presents a shallow minimum at $x=1/2$. Along the other critical lines
the qualitative behavior of the temperature is the same as in 
Fig.~\ref{fig:mf1}. For relatively low $\delta$, the minima of the critical 
temperature located along the critical line arising from the pure species 
are higher than the temperature at the ends of the crescent-shaped line, while
the converse is true for $\delta$ larger than about $0.64$. 

As $\delta$ is increased, the crescent grows towards larger and smaller 
concentrations, until for $\delta=0.65338$ the critical lines meet each other,
resulting in the topology shown in Fig.~\ref{fig:mf3}. When $\delta$ grows
above this value, the former critical line originating from the pure 
components splits: the portions at low and high concentration, where 
the transition is mainly liquid-vapor, join the crescent-shaped line so as 
to form a liquid-vapor critical line ranging from $x=0$ to $x=1$, while 
the part at intermediate concentration, where demixing prevails, remains 
connected to the demixing line at $x=1/2$ and detatches from the liquid-vapor
line, giving a fork-shaped critical locus. The situation just described 
is illustrated in Fig.~\ref{fig:mf4} for $\delta=0.7$. The liquid-vapor 
critical line has just one temperature minimum at $x=1/2$, in agreement with 
the above observation that for a liquid-vapor transition, increasing 
the concentration of the dilute component leads to a decrease of the critical
temperature. The two temperature minima symmetric with with respect to $x=1/2$
are now located along the fork-shaped line. The relative temperature change
along this line is however quite small, as already observed for 
the crescent-shaped line of Fig.~\ref{fig:mf2}. For $\delta$ just above the
value $0.65338$ that marks the boundary between the topology 
of Fig.~\ref{fig:mf2} and that of Fig.~\ref{fig:mf4}, the critical temperature
at the tips of the fork-shaped line is higher than the minimum at $x=1/2$ 
on the liquid-vapor line, while the converse is true at higher $\delta$, 
including the value $\delta=0.7$ to which Fig.~\ref{fig:mf4} refers. A similar
behavior is found when comparing the minimum on the liquid-vapor line with
the temperature at the intersection of the fork with the mixing-demixing line.

If $\delta$ is further increased, the liquid-vapor line becomes more and more
similar to a straight segment, as is to be expected since $\delta=1$ 
corresponds to a one-component fluid, whose critical density is obviously 
independent of the concentration. At the same time, the fork-shaped line
shrinks
and moves to higher density, together with the mixing-demixing line. Strictly
speaking, the fork disappears from the critical lines only 
in the one-component limit $\delta\rightarrow 1$. However, when $\delta$ gets
larger than a value $\delta_{0}$ between $0.75$ and $0.76$, this locus is 
certainly metastable, as the pressure along it is everywhere negative. 
We observe that this is a sufficient condition for metastability, but not 
a necessary one. In fact, according to Ref.~\cite{wilding}, the portion 
of the fork near equimolar concentration has already disappeared from 
the equilibrium phase diagram for $\delta > 0.708$. 
We are then left with two disconnected 
critical lines: the mixing-demixing one, that terminates at an endpoint, and 
the liquid-vapor one at lower density, as shown in Fig.~\ref{fig:mf5} for
$\delta=0.8$. For $\delta\rightarrow 1$, this topology evolves into that 
expected for the one-component fluid, as explained above: the mixing-demixing
line eventually disappears, and the liquid-vapor line becomes a segment 
at constant density. 
    
\section{HRT phase diagram}
\label{sec:results2}
 
As we said in the Introduction, HRT calculations were performed 
for an interaction that consists of a hard-sphere repulsive core and 
an attractive Yukawa tail, whose inverse-range parameter $z$ was set 
to $z=1.8$ for both like and unlike species. This value of $z$ is appropriate
for representing the LJ potential~\cite{henderson} and it has been
widely adopted in the literature. We illustrate our results by considering,
for different values of the parameter $\delta$, several isothermal sections
of the phase diagram, each of which corresponds to a single HRT run. In our
opinion, this gives a clearer picture than the one that would be obtained 
by mapping the phase diagram at given $\delta$ on a single three-dimensional
plot. 
In the following, the temperature will be identified with the 
corresponding reduced quantity $T^{*}=k_{\rm B}T/\epsilon$, and the asterisk
will be omitted.
Fig.~\ref{fig:1a} shows the phase diagram on the density-concentration
plane at four different temperatures for $\delta=0.65$. Because of 
the symmetry of the model considered here, the phase diagram is obviously 
symmetric with respect to the equal-concentration axis $x=1/2$. We already 
pointed out that in the HRT the conditions of thermodynamic equilibrium 
that define the coexistence region are implemented by the theory itself. 
In fact, inside the domains shown in the figures the hessian 
determinant of the Helmholtz free energy 
is identically vanishing. 
We stress that this is essentially different from what
is found in mean field-like approaches, where phase separation is 
marked by the appearance of spinodal surfaces in the $T$-$\rho$-$x$
space, which give spinodal lines upon intersecting with a plane
at constant $T$ like those of the figure. Inside the regions bounded
by these lines, the hessian of the free energy attains unphysically
negative values. In the HRT, on the other hand, the hessian does not
become negative, but it vanishes identically in a region of finite 
measure~\cite{hrt5}. If we consider a curve at given temperature 
and pressure
in the $\rho$-$x$ plane, it is readily seen that along the portion of 
this curve that lies inside the domain where the hessian vanishes, 
the chemical potential of both species are identically constant. 
Therefore, such a domain is indeed the coexistence region 
of the mixture. This clearly appears from Fig.~\ref{fig:1b}, which 
shows that the domains of Fig.~\ref{fig:1a} collapse into lines when
they are plotted in the $P$-$\Delta \mu$ plane, where $P$ 
is the pressure, and $\Delta \mu=\mu_{1}-\mu_{2}$ is the difference between 
the chemical potentials of the two species. Because of the conditions
of thermodynamic equilibrium, the coexistence regions 
in the $T$-$P$-$\Delta \mu$ space appear as ``sheets'' bounded 
by critical lines. Intersecting with a plane at constant $T$ then
yields lines like those of Fig.~\ref{fig:1b}, terminating
at critical points. Each point of these lines corresponds 
to an isobar of the domains of Fig.~\ref{fig:1a}, i.e., 
to a tie-line. At a critical point, the tie-line reduces to a single point. 
We recall that in the $\rho$-$x$ plane a critical point is not in general 
an extremal point of the phase boundary~\cite{rowlinson}, either in $\rho$ 
or in $x$, and 
it cannot be detected by just considering the shape of the $\rho$-$x$ 
coexistence boundaries. Critical points have been marked by dots 
in Fig.~\ref{fig:1a} and in the following figures that show the phase diagram
in the $\rho$-$x$ plane for different values of $\delta$. 

The first panels of Figs.~\ref{fig:1a}, \ref{fig:1b} show the phase
diagram at a temperature somewhat lower than the critical temperature
$T_{c}^{0}\simeq 1.2$ of the pure components. Three distinct coexistence 
domains are present. The two at low density originate from the coexistence 
regions of the pure components, and they both terminate at a critical
point. The high-density region
at $\Delta \mu=0$, which is present also at temperatures above 
$T_{c}^{0}$, involves, at each fixed density, coexistence between two
fluids at concentration $\overline{x}$ and $1-\overline{x}$ 
respectively, and terminates at a mixing-demixing critical point 
at $x=1/2$. Below a certain temperature $T_{t}\simeq 1.06$, however, 
the demixing region
bifurcates into two branches (see the second panel of Fig.~\ref{fig:1b}), 
each of which ends at a critical point with $\Delta \mu \neq 0$, 
$x\neq 1/2$. The former mixing-demixing critical point at $x=1/2$ has now 
become a first-order coexistence boundary, at which a mixed fluid at equal
species concentration coexists with a demixed fluid at higher density similar
to that found above $T_{t}$, consisting of two phases with concentrations
symmetric with respect to $x=1/2$. 
At the temperature $T_{t}$ at which 
the bifurcation develops, the critical line at $x=1/2$ and the two
critical lines generated by the symmetric branches for $T<T_{t}$ meet 
at a tricritical point. These critical lines can be visualized 
as the boundaries of the coexistence ``sheets'' in the $T$-$P$-$\Delta \mu$
space whose projections on the $P$-$\Delta \mu$ plane at constant $T$ 
are shown in Fig.~\ref{fig:1b}. At the tricritical point, the mixed fluid 
at $x=1/2$ and the two phases that constitute the demixed fluid become
critical simultaneously. 
Again, in the $\rho$-$x$ plane the tricritical point does not display any 
special feature that makes it immediately detectable. Inspection 
of the $\rho$-$x$ phase boundary alone does not allow one to tell whether 
the high-density coexistence region displays a single critical point at 
$x=1/2$, two symmetric critical points, or a tricritical point.  
By further lowering the temperature, the high- and
low-density coexistence regions expand and get closer, until at a temperature
$T_{d}\simeq 1.04$ each of the low-density regions meets the high-density 
one at a double critical point. 
The critical lines in the $\rho$-$x$ plane have the same topology 
as in Fig.~\ref{fig:mf1}. In particular, the tricritical point corresponds 
to the intersection of the mixing-demixing critical line at $x=1/2$ 
with the line that spans the concentration axis from $x=0$ to $x=1$. 
The latter actually results from projecting on the $\rho$-$x$ plane the two
critical lines into which the mixing-demixing critical line bifurcates
below $T_{t}$ and those originating from the critical points of the pure
species. The two symmetric double critical points at $T=T_{d}$ where these 
critical lines meet in couples 
correspond to the temperature minima at $x\neq 1/2$ 
located on the critical line of Fig.~\ref{fig:mf1}.  
We observe that, unlike tricritical points, double critical points do not
entail the intersection of topologically distinct critical lines. In fact,
tricritical points are found in the model considered here because of its
special symmetry, but they do not occur in real binary mixtures, while double
critical points are frequently found in real mixtures, including mixtures 
of noble gases such as neon-krypton~\cite{trappeniers1} 
and neon-xenon~\cite{trappeniers2}. Below $T_{d}$, the coexistence domain 
consists of one connected region, without any critical point, as shown
in the third and fourth panels of Fig.~\ref{fig:1b}.
A section of the phase diagram at $x=1/2$ in the $\rho$-$T$ plane is plotted
in Fig.~\ref{fig:eq1}, showing the same behavior as in Fig.~2(d) 
of Ref.~\cite{wilding}.
The HRT and the mean-field critical lines are compared in Fig.~\ref{fig:crit}
for $\delta=0.5$. This value of $\delta$ is small enough to give the same 
topology of the critical lines in HRT and mean-field theory, save for 
an extremely short line at low density and nearly equimolar concentration 
obtained in mean field, which belongs to the metastable regime 
(see the discussion in the previous Section) and has not been reported here.
The figure shows the projections of the critical lines both in the $\rho$-$x$
and in the $x$-$T$ planes; in the latter case, the mixing-demixing line 
at $x=1/2$ has not been shown. For the pure species, it is known from 
the comparison of the mean-field results with accurate simulation data 
for the critical constants of the Yukawa fluid with the same inverse range
$z=1.8$ considered here~\cite{scoza}, that mean-field theory underestimates 
the critical 
density by about $20\%$ and overestimates the critical temperature by 
about $10\%$. Basically the same differences are found by comparing 
the mean-field and the HRT results, as the HRT provides a very good 
determination of the critical point of LJ-like 
fluids~\cite{hrt4,hrtlj,hrtyuk}. Fig.~\ref{fig:crit} shows that similar
discrepancies between mean field and HRT hold also for the critical loci 
of the binary system. We observe that the direction of the order 
parameter is little affected by the inclusion of fluctuations, at least for 
the present case where these do not change the topology of the critical 
lines.    

According to our HRT calculations, 
the value $\delta=0.65$ is actually nearly coincident with the upper limit
of $\delta$ for the topology that we have just described.
The scenario at slightly larger $\delta$ is illustrated 
in Figs.~\ref{fig:2a}, \ref{fig:2b}, which show the phase diagram 
in the $\rho$-$x$ and $P$-$\Delta \mu$ planes for $\delta=0.665$. 
At temperatures $T$ above $1.023$, the phase diagram evolves as before:
on lowering $T$, the high-density coexistence region bifurcates 
at a tricritical point, and each of the resulting branches merges with 
the low-density coexistence regions that originate from the pure species,
so that just below $T=1.03$ there are no critical points left in the phase
diagram. However, at $T=1.023$ two critical points reappear at low density.
In the $P$-$\Delta \mu$ plane, this is marked by the appearance 
of two ``twigs'' that stick out of the low-density coexistence region, 
each terminating at a critical point, as shown in the third panel 
of Fig.~\ref{fig:2b}. By slightly lowering $T$, these twigs grow longer, 
and for $T=T_{d}=1.022$ they meet at a critical double point located at 
$\Delta \mu=0$. At the same time, the low-density lobes of the coexistence
region in the $\rho$-$x$ plane become very elongated, and coalesce at $x=1/2$. 
The topology of the critical lines is the same as in Fig.~\ref{fig:mf2}, where
the crescent-shaped line at low density is that described by the new family
of critical points. We remark that, according to mean-field theory, this line
is extremely shallow with respect to the temperature. This is confirmed 
by the HRT results just reported, which give a relative variation 
of the critical temperature of about $0.1\%$. Below $T_{d}$ (see the fourth 
panel of Fig.~\ref{fig:2a}), the phase diagram presents an ``island'' 
of homogeneous, mixed fluid surrounded by a ``sea'' of phase-separated fluid.
A section of the coexistence region at $x=1/2$ shows that in this regime 
we have two kinds of phase equilibria at equal species concentration: on the
one hand, below the tricritical temperature $T_{t}$ a demixed fluid at high 
density coexists with a mixed one at intermediate density as before. 
On the other hand, this mixed fluid coexists, at slightly lower density, 
with another low-density mixed fluid. The latter phase equilibrium terminates
for $T=T_{d}$ at a vapor-liquid critical point. Therefore, the phase diagram
at $x=1/2$ exhibits both a critical and a tricritical point. 
On further lowering the temperature below $T_{d}$, the island of mixed fluid
shown in Fig.~\ref{fig:2a} becomes smaller and smaller, until it is eventually
swallowed by the coexistence region. Correspondingly, in the $\rho$-$T$ plane
the two phase boundaries at $x=1/2$ meet at a triple point, where  
the two mixed fluids at different densities coexist with 
the demixed fluid~\cite{triple}. Below the temperature 
of the triple point, the demixed fluid 
at high density coexists with the mixed one at low density. 
The phase diagram at equal species concentration
is similar to Fig.~2(c) of Ref.~\cite{wilding} 
and to that shown below in Fig~\ref{fig:eq2} for the case $\delta=0.67$. 
From the picture given above
it is clear that the low-density lobes of the coexistence region meet 
at a temperature lower than that at which each lobe meets the high-density
region. This in turn is lower than the temperature at which the tricritical
point appears. Therefore, for the topology just discussed the liquid-vapor
critical temperature $T_{d}$ is always lower than the tricritical temperature
$T_{t}$. 

A different situation arises as $\delta$ grows above $0.67$. 
Figs.~\ref{fig:3a}, \ref{fig:3b} depict the evolution of the phase diagram 
for $\delta=0.68$. As before, a tricritical point is present at $x=1/2$,
as shown by the bifurcation of the high-density coexistence region. 
On the other hand, on lowering the temperature the low-density branches of
the coexistence region do not merge with the high-density one as before,
but instead meet each other at a critical double point between $T=1.03$ 
and $T=1.025$. In this regime the isothermal sections of the phase diagram 
in the $\rho$-$x$ plane consist of two disconnected domains, with that 
at lower density spanning the whole density-concentration axis from $x=0$ 
to $x=1$. At about $T=1.025$, two twigs sprout out of the low-density 
coexistence domain in the $P$-$\Delta \mu$ plane. In such a situation
this domain presents two symmetric critical points, corresponding 
to the ends of the twigs. On further lowering $T$, the twigs grow longer, 
and for $T$ between $1.015$  and $1.01$ each of them meets a branch 
of the high-density coexistence region at a critical double point. 
The low- and high-density coexistence domains then join at two points 
symmetric with respect to $x=1/2$, leaving between them an island of one-phase 
fluid (see Fig.~\ref{fig:3a}), that eventually disappears at low temperature. 
This scenario is somehow the converse of that previously described for 
$\delta=0.665$, where the high-density coexistence domain merges with 
the low-density ones originating from the pure components, and the one-phase
island inside the coexistence region is formed because of the twigs meeting
each other at $x=1/2$. The topology of the critical lines for the case just
discussed is that of Fig.~\ref{fig:mf4}: this time the critical line that goes 
from $x=0$ to $x=1$ is not connected to the mixing-demixing critical line, 
and it has a temperature minimum at $x=1/2$, corresponding to the coalescence
of the low-density coexistence regions into one connected domain. 
The mixing-demixing critical line intersects at the tricritical point  
two symmetric critical lines as before, which however do not meet those 
originating from the pure species. Instead, they merge with the lines 
described by the family of critical points that sprout out of the low-density 
coexistence region, resulting in the fork-shaped line of Fig.~\ref{fig:mf4}. 
This presents two symmetric temperature minima at the value of $T$ at which 
the critical lines meet each other. As already observed for 
the crescent-shaped line of Fig.~\ref{fig:mf2}, the temperature along the fork
is actually almost constant, both according to mean-field theory and HRT, 
so that the minima are extremely shallow. The topology of the phase diagram 
at equimolar concentration is similar to the previous one illustrated
in Fig.~\ref{fig:eq2} for $\delta=0.67$. In particular,
in a certain temperature interval including that where the two-phase region
encloses a one-phase domain, there are two different phase equilibria 
at $x=1/2$. For the coupling parameter $\delta=0.68$ considered here, 
the liquid-vapor critical temperature $T_{d}$, which is the temperature 
minimum along the liquid-vapor critical line, is smaller than the tricritical
temperature $T_{t}$, as for $\delta=0.665$. However, in the present topology
where, on lowering $T$, the low-density coexistence domains merge 
with each other before they merge with the high-density one, nothing prevents
the liquid-vapor critical point at $x=1/2$ from occurring at a higher 
temperature than the tricritical point. In the $P$-$\Delta \mu$ plane, this
corresponds to the high-density coexistence region bifurcating 
at a temperature lower than that at which the low-density branches meet. 
Such a scenario in fact comes along starting from about $\delta=0.7$. 
The phase diagram at $x=1/2$ for this value of $\delta$
is shown in Fig.~\ref{fig:eq3}, and
is the same at that of Fig.~2(b) of Ref.~\cite{wilding}. 

The phase behavior of the system for $\delta$ intermediate between the values
$0.665$ and $0.68$ discussed above is worth being considered in more detail.
Figs.~\ref{fig:4a}, \ref{fig:4b} show the phase diagram for $\delta=0.67$. 
On lowering the temperature, the low-density coexistence domains coalesce 
at about $T=1.0235$, and at a slightly lower temperature around $T=1.023$ 
the resulting region coalesce with the high-density coexistence domain.
The phase diagram in the $\rho$-$x$ plane is then similar to that already 
shown for $\delta=0.68$, except that here the low- and high-density domains
meet at nearly the same temperature. However, if one considers temperatures
just above $T=1.0235$ it appears that, unlike what found for the $\delta$
studied above, the phase diagram displays six critical points, two for each
of the disconnected domains that make up the coexisting region. This is 
clearly shown in the $P$-$\Delta \mu$ plane of Fig.~\ref{fig:4b} 
for $T=1.024$. In the present case it is not obvious to tell whether 
the topology of the critical lines in the $\rho$-$x$ plane is that 
of Fig.~\ref{fig:mf2} or of Fig.~\ref{fig:mf4}, because it is difficult 
to ascertain which is the part of the coexistence region 
in the $P$-$\Delta \mu$ plane that is sprouting from the other, 
and consequently whether the high-density branches of the coexistence domain
are bound to meet those originating from the pure species 
as for $\delta=0.655$, or the twigs as for $\delta=0.68$. We are therefore 
very close to the boundary between the two topologies of the critical loci, 
that corresponds to the scenario of Fig.~\ref{fig:mf3}. This is marked 
by the presence of two tricritical points at non-equimolar concentration 
besides that at $x=1/2$. In such a situation the growth of a secondary 
structure (the twigs) from an already existing one (the main branch) is 
replaced by a bifurcation where both branches stem out of the tricritical
point at the same temperature. For continuity reasons, we expect that 
in a very narrow range of $\delta$ values that includes $0.67$ 
and is contained in the interval $0.665<\delta<0.68$, the configuration with
six critical points shown in Fig.~\ref{fig:4b} can be found for the topology
of the critical lines of both Fig.~\ref{fig:mf2} and Fig.~\ref{fig:mf4}. 
If one considers the critical lines of Fig.~\ref{fig:mf2}, this occurs when
the critical temperature at the ends of the low-density crescent-shaped line
is higher than the minima located along the line originating from the pure
species. In the case of Fig.~\ref{fig:mf4}, the requirement is that the local
temperature minimum at $x=1/2$ along the liquid-vapor line must be lower 
not only than the tricritical temperature, but also than that of the critical 
points located at the tips of the fork-shaped line. It may also be worth 
pointing out that, strictly speaking, the watershed between
the density-concentration phase diagrams of Fig.~\ref{fig:2a} 
and Fig.~\ref{fig:3a}, that represent the two different ways by which a domain
of one-phase fluid can be enclosed into the coexistence region, has not 
to coincide with the boundary between the critical lines just discussed, 
although the two are expected to occur for very similar values of $\delta$.
In fact, according to mean-field theory, when the critical lines have 
the topology of Fig.~\ref{fig:mf3}, the coexistence region in the $\rho$-$x$
plane still looks like that of Fig.~\ref{fig:2a}. This implies that there
is an extremly narrow range of $\delta$ where the topology of the critical
lines is as in Fig.~\ref{fig:mf4}, and the topology of the phase diagram 
in the $\rho$-$x$ plane is as in Fig.~\ref{fig:2a}. In the $P$-$\Delta \mu$ 
plane such a regime is marked by the twigs joining the bifurcation 
of the high-density coexistence region at a temperature higher than that 
at which the low-density branches meet each other. We have not checked whether
this scenario comes along also in the HRT, or instead it is replaced 
by the converse one, where the critical lines have the topology 
of Fig.~\ref{fig:mf2}, and the phase diagram in the $\rho$-$x$ plane 
has the topology of Fig.~\ref{fig:3a}. 

Further information on the nature of the phase equilibria is obtained 
by considering the tie-lines, i.e., the lines in the $\rho$-$x$ plane 
that connect the phases at coexistence at a certain temperature
and pressure. A few tie-lines of the mixture with $\delta=0.67$ are shown
in Fig.~\ref{fig:tie} for $T=1.024$ (left panel) and $T=1.023$ (right panel), 
corresponding respectively 
to the second and fourth panel of Fig.~\ref{fig:4a}. We have chosen to report
the tie-lines for the case $\delta=0.67$ because of its particularly rich 
phase diagram, several features of which are separately found also for 
different values of $\delta$. In both panels of Fig.~\ref{fig:tie}, 
the high-density portion of the coexistence region is characterized 
by the presence of a demixed fluid which, as said above, consists 
of two phases at the same density and concentrations symmetric with respect
to the equal-concentration axis. As a consequence, the average density 
coincides with that of the coexisting phases irrespective of their relative
amount in the demixed fluid, resulting in strictly vertical tie-lines. 
Since both panels refer to temperatures below the tricritical temperature 
$T_{t}$, the phase equilibrium just described does not extend down 
to the left boundary of the high-density coexistence domain. At a certain
density $\rho_{\rm D}$, the two coexisting fluids at symmetric concentrations
$\overline{x}$, $1-\overline{x}$ coexist in turn with a fluid at a lower density
$\rho_{\rm M}$ and equimolar concentration. The densities $\rho_{\rm M}$, 
$\rho_{\rm D}$ are the boundaries of the first-order coexistence domain 
that separates the mixed-fluid and demixed-fluid regions of Fig.~\ref{fig:eq2}
for $T<T_{t}$. In the $\rho$-$x$ plane, this three-phase equilibrium takes 
place in a triangle-like domain bounded by three tie-lines meeting in couples
at the points $(\rho_{\rm D},\overline{x})$, $(\rho_{\rm D},1-\overline{x})$,
$(\rho_{\rm M}, 1/2)$. These lie very close to (but do not exactly 
coincide with)
the two oblique lines and the leftmost vertical line shown in the high-density
region of Fig.~\ref{fig:tie}.  
The two symmetric portions of the high-density region 
of the left panel that lie above and below the three-phase domain correspond
in the $P$-$\Delta \mu$ plane to the branches of the bifurcation that stems
from the line at $\Delta \mu=0$ (see Fig.~\ref{fig:4b}). Each of them presents
coexistence between two phases which differ both in density and concentration,
terminating at a critical point where the tie-line reduces to a single point.
In the right panel the critical points are absent, as the high- and 
low-density coexistence regions have merged into one connected domain. 
The left panel also shows the tie-lines shrinking at the four critical points
located on the two symmetric lobes of the low-density coexistence region. 
An interesting feature that appears from the arrangement of the tie-lines 
in this region is the presence of two more domains of three-phase coexistence, 
such that the coexisting phases have concentrations that lie on the same side
of the $x=1/2$ axis. At temperatures at which
the coexistence region is connected and encloses an island of homogeneous 
fluid as in the right panel, this sort of three-phase equilibrium could 
be expected just on the
basis of the fact that different tie-lines cannot intersect each other.
However, the presence in the low-density region of the two critical points
associated with the twigs in the $P$-$\Delta \mu$ plane implies that 
three-phase coexistence is observed in this region even at temperatures  
at which it has not yet merged with the high-density coexistence domain. 
We also observe that in the neighborhood of the $x=1/2$ axis where the two 
lobes of the low-density region coalesce, the tie-lines are nearly horizontal,
meaning that the concentration of the coexisting phases are very similar. 
At exactly $x=1/2$ azeotropy occurs, i.e., one finds a purely liquid-vapor
transition between two phases at equal concentration, as already observed 
above in connection with the phase diagram at $x=1/2$ in the $\rho$-$T$ plane.
The critical point topping the liquid-vapor coexistence curve at equimolar
concentration in Figs.~\ref{fig:eq2}, \ref{fig:eq3} is therefore an azeotropic
critical point.      
      
As $\delta$ increases above $0.7$, the curvature of both the low- and
high-density coexistence boundaries in the $\rho$-$x$ plane rapidly 
decreases, and the pressure at which the bifurcation 
of the high-density coexistence region takes place gets closer and
closer to that of the coexisting vapor and liquid phases 
on the low-density region at $x=1/2$. As a consequence, the one-phase
island bounded by the coexistence region becomes smaller and involves
a fluid at nearly equimolar concentration. One can then ask whether 
this feature of the phase diagram will disappear at a certain 
$\delta_{0}<1$, or instead only in the limit $\delta \rightarrow 1$. 
The former case corresponds to a situation where for 
$\delta > \delta_{0}$ the high-density coexistence region does not
undergo any bifurcation in the $P$-$\Delta \mu$ plane, but it always
presents a mixing-demixing critical point at $x=1/2$, $\Delta \mu=0$. 
At low enough temperature, this critical point meets the low-density
coexistence region at a first-order phase boundary. The topology of
the critical lines in the $\rho$-$x$ plane is that of  
Fig.~\ref{fig:mf5}: the mixing-demixing line terminates at a critical
enpoint where the low- and high-density coexistence regions meet. 
The phase diagram at $x=1/2$ corresponding to this scenario is that
of Fig.~2(a) of Ref.~\cite{wilding}. As before, the equimolar mixture
displays both a liquid-vapor transition that occurs below a certain 
critical temperature and involves two mixed fluids of different 
densities, and a mixing-demixing transition at high density. In this
case, however, the transition between the mixed liquid and 
the demixed fluid is always second-order down to the temperature 
of the endpoint, below which the demixed fluid coexists with a mixed
vapor of much lower density. The situation where 
the contact between the 
low- and the high-density coexistence region always occurs at two
distinct points symmetric with respect to $x=1/2$ until $\delta$ gets
equal to $1$ is instead consistent with 
the phase diagram remaining qualitatively similar 
to that found for $\delta=0.7$ with
critical lines as in Fig.~\ref{fig:mf4}, albeit the 
fork-shaped line will become vanishingly small as $\delta$ approaches
$1$ and the mixing-demixing line will be pushed to higher density. 
The phase diagram of the equimolar mixture in the $\rho$-$T$ plane 
will qualitatively look as that of Fig.~\ref{fig:eq3}. 
As a consequence, as long as one has $\delta < 1$ there will be 
a small temperature interval between the triple point and 
the tricritical point, where the transition between the mixed 
and the demixed liquid is first order. This is the scenario advocated
in Ref.~\cite{forstmann} on the basis of modified hypernetted chain 
(MHNC) calculations on a symmetric LJ mixture, while the
simulations performed in Ref.~\cite{wilding} on a square-well system
supported the existence of a critical endpoint at high enough 
$\delta$. Some calculations that we previously performed on 
a hard-sphere LJ mixture using a $100 \times 100$
density-concentration grid also suggested that for $\delta=0.8$ 
the phase diagram of the system presents a critical endpoint. 
However, the present results for a hard-core Yukawa potential 
shown in Figs.~\ref{fig:5a}, \ref{fig:5b} indicate that for $\delta=0.8$,  
the mixture most likely does not have a critical 
endpoint: in fact, in a certain narrow temperature range 
the coexistence region in the $\rho$-$x$ plane still encloses 
a domain of one-phase fluid with concentration varying in 
a small interval around $x=1/2$. Since the inverse range $z=1.8$ of 
the Yukawa potential considered here gives a fair representation of
the LJ interaction~\cite{henderson}, the difference 
between this behavior and that reported in Ref.~\cite{hrt3} 
most probably does not depend on some intrinsic difference between 
the two interactions, but just on the fact that the higher resolution
entailed by the $150 \times 150$ grid used in this work allows one 
to study the contact between the low- and high-density coexistence
regions with better accuracy than in Ref.~\cite{hrt3}, thereby 
uncovering the presence of a small one-phase domain. 
We should observe that the case $\delta=0.8$ 
actually lies at the limit of resolution even for the larger grid
employed here: the homogeneous domain inside the coexistence region
in the $\rho$-$x$ plane extends only one gridpoint in the 
$\rho$-direction, while the phase diagram in the $P$-$\Delta \mu$ 
plane does not present any of the features associated with 
the occurrence of such a domain on the scale of the figure, 
so that it cannot be clearly distinguished from the kind of diagram one would
expect in the presence of a critical endpoint. 
In particular, there is no sign of the bifurcation of the high-density 
coexistence region in the $P$-$\Delta \mu$ plane that we expect if 
the low- and high-density regions have to meet at two points 
of non-equimolar concentration according to the scenario of Fig.~\ref{fig:3b}. 
This most probably depends on the fact that, 
for the relatively high value of $\delta$ considered here, the densities 
of the mixed and demixed liquids at coexistence are so close, that 
their difference is comparable to the mesh size. As a consequence, 
the differences in the pressures and chemical potentials 
of neighboring but non-coexisting points will also become comparable 
to the numerical errors.  
A little numerical noise in the $P$-$\Delta \mu$ plane due to the finite
mesh used in the calculation is in fact always present, as shown by close
inspection of the relevant figures, but only at high $\delta$ does this 
become a hindrance for a clear description of the phase behavior. 
We did not pursue
any investigation for $\delta > 0.8$, because {\em a fortiori} in such 
a regime 
the resolution allowed by the grid used here would not enable us to
discriminate between the occurrence of a critical endpoint, or 
an extremely small region of homogeneous fluid that undergoes 
a first-order demixing transition on slightly increasing the density. 
Therefore, we are not in a position to say whether a critical 
endpoint at $x=1/2$ will eventually appear near $\delta=1$, or instead a very
weak first-order transition will survive up to $\delta=1$. Another
issue that is difficult to elucidate within the resolution of the
present calculation concerns the fate of the twigs discussed before
in connection with the phase diagram topology of Figs.~\ref{fig:3a},
\ref{fig:3b}. As noted above, even at temperatures at which the
low-density coexistence regions that originate from the pure species
have coalesced into one connected domain, the latter can still show
two symmetric critical points. When this is the case, the low-density
coexistence region in the $P$-$\Delta \mu$ plane presents two small
twigs, each of them ending at a critical point. 
These twigs could either disappear at high enough $\delta$,
or persist for $\delta$ arbitrarily close to $1$. 
If a critical
endpoint at $x=1/2$ never appears and the topology of the critical lines 
in the $\rho$-$x$ plane is that of Fig.~\ref{fig:mf4} until $\delta$ gets
equal to unity, 
the latter case corresponds to a situation where 
the low- and high-density coexistence regions will always meet at two double
critical points. As discussed above, mean-field theory indeed gives two 
double critical points located along the fork-shaped line, but for $\delta$ 
close to $1$ they are certainly metastable since they correspond to negative 
pressure. If instead the twigs disappear before $\delta$ reaches $1$, there
will be a value of $\delta$ above which the critical points 
of the high-density coexistence region meet the low-density region at two
points of first-order coexistence. As a consequence, the contact takes place
at two critical endpoints with concentrations symmetric with respect to
$x=1/2$ rather than at two double critical points. This is the situation 
considered in Ref.~\cite{forstmann}. In such a case the critical lines 
in the $\rho$-$x$ plane have still the same topology 
as in Fig.~\ref{fig:mf4}, except that the temperature minima along 
the fork-shaped line have now to coincide with the tips of the fork, which
correspond to the two endpoints.  
It can also be worthwhile observing that in principle the twigs are 
not incompatible with a critical endpoint at equimolar concentration. In fact,
this could even give rise to a scenario where the presence of the endpoint
is consistent with the high- and low-density coexistence regions meeting 
at non-equimolar concentration. Specifically, if the critical points located
at the ends of the twigs meet the high-density coexistence region 
at two symmetric critical endpoints with $x\neq 1/2$, nothing prevents 
the point 
at equimolar concentration of this region from remaining critical, until it 
also meets the low-density region at a critical endpoint. We remark that such
a possibility hinges on the presence of the twigs and the related critical 
points on the low-density coexistence region. In the converse situation 
where the critical endpoints at $x\neq 1/2$ result from the contact between 
two points of first-order coexistence on the low-density domain 
and two critical points on the high-density one, 
the requirement that the point at $x=1/2$ be critical is
untenable, as pointed out 
in Ref.~\cite{forstmann}. The scenario just depicted is somewhat suggestive,
as it could account for both the presence of a critical endpoint at $x=1/2$
above a certain value of $\delta$ reported in simulation 
studies~\cite{wilding}, 
and the failure to observe the low- and high-density coexistence regions 
coalescing at $x=1/2$ found in the present investigation as well as 
in~\cite{forstmann}. However, we must point out that at the present stage
we do not have any solid evidence that this possibility does actually occur, 
so that we must regard it as a purely speculative conjecture.

\section{Conclusions}
\label{sec:conclusions} 
      
We have used the HRT to perform an investigation of the phase diagram 
of symmetric binary mixtures as a function of the unlike-to-like interaction
ratio $\delta$. The microscopic interaction adopted consisted 
of a hard-core repulsion plus an attractive Yukawa tail potential with inverse
range $z=1.8$. Such a potential has been used many times in liquid-state 
theory to describe a simple LJ-like fluid. For each value of 
$\delta$ considered, results for the coexistence regions were obtained 
on the whole density-concentration plane at several temperatures. 
The resulting phase portrait was related both to the different topologies
of the mean-field critical lines that come along as $\delta$ is varied
in the interval $0<\delta<1$, and to the behavior predicted by mean-field
theory and simulation results~\cite{wilding} for a certain special class 
of the systems considered here, namely equimolar mixtures at molar fraction
$x=1/2$. A feature of the HRT that is particularly useful is that the domains
of coexisting phases are straightforwardly obtained by the theory as the loci
where the conditions of thermodynamic equilibrium between different phases
are satisfied, without any need of enforcing them {\em a posteriori}. 
Fulfillment of these conditions is shown by the collapse of the isothermal 
sections of the coexistence regions on lines of the $P$-$\Delta \mu$ plane,
$P$ being the pressure, and $\Delta \mu$ the difference between the chemical
potentials of the two species. This property makes also easy to identify 
the critical points exhibited by the phase diagram at a certain temperature. 

According to the results previously obtained by mean-field theory 
and simulations, the phase diagram of symmetric mixtures at equimolar 
concentration is characterized by three different regimes, depending 
on the value of $\delta$: at low $\delta$ ($\delta<0.605$ according 
to mean-field theory~\cite{wilding}) a mixing-demixing critical line joins
a mixing-demixing coexistence curve at a critical point, thereby generating
a tricritical point. For $\delta$ closer to $1$ ($\delta > 0.708$ 
in mean field~\cite{wilding}), the mixing-demixing critical line joins 
a liquid-vapor coexistence curve at a critical endpoint. Finally, in a narrow
interval in $\delta$ intermediate between the above regimes, the mixture 
shows both a liquid-vapor coexistence curve and a mixing-demixing one 
at higher densities, topped respectively by a critical and a tricritical 
point. Our investigation clearly shows the tricritical point regime as well
as the intermediate one. The latter is predicted to occur starting from
about $\delta > 0.65$, in agreement with the simulation results for 
a square-well mixture~\cite{wilding}, but it lingers on for larger values 
of $\delta$ than those given by simulation: according to simulation, the 
endpoint regime is reached for $\delta > 0.68$, while according to HRT 
$\delta=0.7$ is still in the ``transition'' or intermediate 
regime, as shown in Fig.~\ref{fig:eq3}. As a matter of fact,  
we did not find any clear
evidence of the endpoint regime, since our calculations indicate 
that the mixture is likely to be in the 
intermediate regime for $\delta$ as high as $0.8$.
However, on increasing $\delta$, the transition region quickly moves to high
density, and the finite resolution allowed by the $150\times 150$ grid used 
here becomes insufficient to fully uncover the topology of the phase diagram
in this parameter range. For this reason we did not investigate the behavior
of the model for $\delta>0.8$.
The results obtained
in this work are in qualitative agreement with a study based on the MHNC
integral equation for a LJ mixture~\cite{forstmann}, according
to which no critical endpoint at equimolar concentration is present 
up to at least $\delta=0.81$. 
Before drawing any definite conclusion about the resilience 
of the intermediate topology in the theoretical HRT and MHNC results compared 
with
the simulations and the mean-field approximation, the role of the specific
interaction adopted should be elucidated. In fact, 
the HRT and MHNC calculations were performed on a HCY 
and on a LJ potential respectively, while in the simulations 
a hard-sphere plus square-well was used. 
As observed in Sec.~\ref{sec:introduction}, the HCY potential was also 
employed in other
investigations based on the MSA~\cite{kahl1}, the ORPA~\cite{kahl2}, 
and the SCOZA~\cite{kahl3} theories, all of which yield for the phase diagram
the same qualitative picture found in mean-field theory. 
However, we do not see any general reason why the independency 
of the interaction profile which is intrinsic to the mean-field phase behavior
should always hold also for more sophisticated approaches, specifically for 
the HRT.   
Moreover, one can also imagine a scenario where the existence of a critical
endpoint at equimolar concentration is not prevented by the behavior observed 
here for the phase diagram in the $\rho$-$x$ plane, namely the fact that,  
even for relatively high $\delta$, the low- and high-density coexistence 
regions meet first at two points symmetric with respect to $x=1/2$ rather 
than at $x=1/2$. However, the present limits in resolution of our numerical
calculation do not allow us to say whether this conjecture is actually 
relevant for the system studied. 

An interesting issue that was considered here is how the phase portraits 
corresponding to the topologies mentioned above look like, if one moves off
the plane of equimolar concentration. In particular, the intermediate regime
is marked by the presence, in a certain temperature interval, of a phase
coexistence domain with a hole of homogeneous fluid inside it. This comes 
along according to two distinct scenarios that can occur as the temperature 
is lowered: for $0.65 < \delta < 0.67$, the two low-density coexistence 
regions originating from the pure species meet the high-density coexistence
region associated with the mixing-demixing transition, and subsequentely 
they meet each other, leaving a domain of mixed fluid enclosed inside the 
two-phase region. For $\delta>0.67$, first the two low-density coexistence
regions meet each other, and then the resulting connected domain meets 
the high-density coexistence region at two points symmetric with respect 
to concentration $x=1/2$, leaving some one-phase fluid in between. 
In the former case, the temperature $T_{c}$ of the liquid-vapor critical point
at $x=1/2$ is always lower than the tricritical temperature $T_{t}$, while 
in the latter case $T_{c}$ can be either lower or higher than $T_{t}$, 
depending on $\delta$. The present calculation gives $T_{t}<T_{c}$ starting
from about $\delta=0.7$. 

Because of the existence of both a liquid-vapor and a demixing 
transition, the critical loci show both a line that spans the concentration
axis and connects the critical points of the pure components, 
and a mixing-demixing line at $x=1/2$. The tricritical point topology 
corresponds to a situation where these lines are connected, and the character
of the transition changes continuously from liquid-vapor at $x=0$ or $x=1$
to mixing-demixing at $x=1/2$. In the endpoint regime, which as said above
was not observed in the present investigation, the critical lines are instead
disconnected, and the transition along the line that connects the critical 
points of the pure species is essentially of liquid-vapor type. 
The intermediate regime between these topologies is characterized 
by the presence of a further critical line at concentrations ranging 
in a certain interval centered at $x=1/2$, which was referred to above as 
either the ``crescent'' or the ``fork'' line. The number of critical points
that are found for a certain isothermal section of the phase diagram depends
on the relative location of the temperature extrema along the critical lines. 
The behavior found by the present HRT calculation agrees qualitatively with 
that given by mean-field theory. In particular, the system is predicted 
to have up to six critical points at a certain temperature. According to HRT,
this occurs for a very narrow range of $\delta$ values contained 
in the interval $0.665<\delta<0.68$. At the boundary between the ``crescent''
and the ``fork'' topologies of the critical lines, the system has two 
tricritical points at symmetric non-equimolar concentrations besides that 
at concentration $x=1/2$. In HRT, this particular topology 
occurs for a value of $\delta$ close to $0.67$, to be compared with 
the mean-field result $\delta=0.65338$. In both cases, these values 
lie very near the boundary between
the two different types of one-phase ``holes'' in the phase coexistence region
described above. 
When $\delta$ is such that mean-field theory and HRT predict the same 
qualitative topology of the phase diagram, the quantitative discrepancy 
between the critical loci given by the two approaches is similar to that found
for the pure species.

This investigation shows that symmetric mixtures, despite their conceptual 
simplicity, exhibit a very rich phase behavior.  
Like real mixtures, these systems have both liquid-vapor and mixing-demixing
transitions. However, while in real mixtures the liquid-vapor 
and the mixing-demixing regimes generally correspond to states that differ 
widely in density and pressure, in symmetric mixtures these transitions
can instead be located in the same region of the thermodynamic space. 
As a consequence, they tend to compete with each other, so that even a small 
variation in the relative strength of the interactions expressed 
by the parameter $\delta$ is sufficient to bring about significant
qualitative changes in the phase diagram.       
The results presented here are also relevant for Ising ferrofluids in the
presence of a magnetic field. More generally, they show that HRT is capable 
of providing a comprehensive 
description and resolving even subtle features of the phase behavior 
of the model. This ability could prove useful to systematically
study beyond the mean-field level also the phase diagram of more realistic, 
non-symmetric model mixtures that depend one more than just one parameter.

\begin{figure}
\caption{Density-concentration projection of the mean-field critical
lines of a symmetric mixture with interaction parameter 
$\delta=\epsilon_{12}/\epsilon_{11}=0.4$. The total density $\rho$ is the
sum of the densities $\rho_{1}$, $\rho_{2}$ of the components, 
and the concentration $x$ is defined as $\rho_{1}/\rho$.  
The open dots mark the locations of the local minima in the critical 
temperature.
The arrows indicate the direction of the order parameter (see text).}
\label{fig:mf1}
\end{figure}

\begin{figure}
\caption{Same as Fig.~\protect\ref{fig:mf1} for $\delta=0.65$.} 
\label{fig:mf2}
\end{figure}

\begin{figure}
\caption{Same as Fig.~\protect\ref{fig:mf1} for $\delta=0.65338$.}
\label{fig:mf3}
\end{figure}

\begin{figure}
\caption{Same as Fig.~\protect\ref{fig:mf1} for $\delta=0.7$.}
\label{fig:mf4}
\end{figure}

\begin{figure}
\caption{Same as Fig.~\protect\ref{fig:mf1} for $\delta=0.8$.}
\label{fig:mf5}
\end{figure}

\begin{figure}
\caption{Isothermal sections of the coexistence region of a symmetric 
HCY mixture  
in the $\rho$-$x$ plane according to the HRT. The inverse interaction range
is equal to $z=1.8$ and the interaction ratio is equal to $\delta=0.65$.
The dots mark the locations of the critical points.} 
\label{fig:1a}
\end{figure}

\begin{figure}
\caption{Same as Fig.~\protect\ref{fig:1a} in the pressure-chemical potential
plane. $\Delta \mu=\mu_{1}-\mu_{2}$ is the difference between the chemical 
potentials of the components. Note how the points of the coexistence regions 
shown in Fig.~\protect\ref{fig:1a} collapse into lines as a consequence 
of the conditions of thermodynamic equlibrium.} 
\label{fig:1b}
\end{figure}

\begin{figure}
\caption{Phase diagram in the $\rho$-$T$ plane of the HCY 
mixture with $\delta=0.65$ for the special case of equimolar 
concentration $x=1/2$. Open and full dots denote respectively the 
$\lambda$-line and the first-order phase boundary.}  
\label{fig:eq1}
\end{figure}

\begin{figure}
\caption{Critical lines in the $\rho$-$x$ plane (upper panel) and $x$-$T$ 
plane (lower panel) of the HCY mixture with $\delta=0.5$. 
Dotted lines: mean-field theory. Full dots: HRT. In the upper panel, the solid 
line is a guide for the eye obtained by smoothly interpolating between 
the dots, and the arrows indicate the direction of the order parameter. 
The order
parameter along the mixing-demixing critical line at $x=1/2$ is not shown here 
for clarity and is parallel to the $x$-axis according to both mean field 
and HRT. 
In the lower panel, the mixing-demixing critical line is not shown.}  
\label{fig:crit}
\end{figure}

\begin{figure}
\caption{Same as Fig.~\protect\ref{fig:1a} for $\delta=0.665$.}
\label{fig:2a}
\end{figure}

\begin{figure}
\caption{Same as Fig.~\protect\ref{fig:1b} for $\delta=0.665$.}
\label{fig:2b}
\end{figure}

\begin{figure}
\caption{Same as Fig.~\protect\ref{fig:eq1} for $\delta=0.67$.
The letters V, L denote respectively the vapor and mixed liquid phase.}
\label{fig:eq2}
\end{figure}

\begin{figure} 
\caption{Same as Fig.~\protect\ref{fig:1a} for $\delta=0.68$.}
\label{fig:3a}
\end{figure} 

\begin{figure}
\caption{Same as Fig.~\protect\ref{fig:1b} for $\delta=0.68$.}
\label{fig:3b}
\end{figure}

\begin{figure}
\caption{Same as Fig.~\protect\ref{fig:eq2} for $\delta=0.7$.} 
\label{fig:eq3}
\end{figure}

\begin{figure}
\caption{Same as Fig.~\protect\ref{fig:1a} for $\delta=0.67$.}
\label{fig:4a}
\end{figure}

\begin{figure}
\caption{Same as Fig.~\protect\ref{fig:1b} for $\delta=0.67$.}
\label{fig:4b}
\end{figure}

\begin{figure}
\caption{Coexistence region in the $\rho$-$x$ plane for $\delta=0.67$ 
at $T=1.024$ (left panel) and $T=1.023$ (right panel), corresponding  
to the second and fourth panels of Fig.~\protect\ref{fig:4a}, showing 
a few tie-lines connecting phases at coexistence. The different shades 
of gray give a measure of pressure 
(black: low pressure; white: high pressure).}
\label{fig:tie}
\end{figure}  

\begin{figure}
\caption{Same as Fig.~\protect\ref{fig:1a} for $\delta=0.8$. The critical 
points in the second panel are not shown, because for this value 
of $\delta$ the resolution of our numerical calculation does not allow us
to locate them.}
\label{fig:5a}
\end{figure}

\begin{figure}
\caption{Same as Fig.~\protect\ref{fig:1b} for $\delta=0.8$. In the first
panel, the high-density coexistence region at $\Delta \mu=0$ does not appear 
as its pressures are outside the scale of the figure.}
\label{fig:5b}
\end{figure}


\begin{thebibliography}{99}
\bibitem{triple} Strictly speaking, at the triple point there are four 
coexisting phases, since the demixed fluid consists of two homogeneous fluids 
at concentrations symmetric with respect to $x=1/2$.
\end{thebibliography}
\end{document}